\documentstyle[prl,twocolumn,aps,tighten,floats,epsfig]{revtex}

\begin{document}
\draft
\twocolumn[\hsize\textwidth\columnwidth\hsize\csname @twocolumnfalse\endcsname
\title{Theory of Coexisting Charge and Spin-Density Waves in (TMTTF)$_2$Br,
(TMTSF)$_2$PF$_6$ and $\alpha$-(BEDT-TTF)$_2$MHg(SCN)$_4$}

\author{S. Mazumdar and S. Ramasesha$^*$}

\address{Department of Physics, University of Arizona, Tucson, AZ 85721}

\author{R. Torsten Clay and David K. Campbell}

\address{Department of Physics, University of Illinois at 
Urbana-Champaign, Urbana, IL 61801}
\maketitle
\begin{abstract}
Recent experiments indicate that the spin-density waves (SDWs) in
(TMTTF)$_2$Br, (TMTSF)$_2$PF$_6$ and $\alpha$-(BEDT-TTF)$_2$MHg(SCN)$_4$ 
are 
highly
unconventional and coexist with charge-density waves (CDWs). We present a
microscopic theory of this unusual CDW-SDW coexistence. A complete
understanding requires the explicit inclusion of strong Coulomb 
interactions, lattice discreteness, 
the anisotropic two-dimensional nature of the lattice, and the correct 
bandfilling
within the starting Hamiltonian.
\end{abstract}
]
\noindent PACS indices: 71.30.+h, 71.45.Lr, 75.30.Fv, 74.70.Kn
\narrowtext
Organic charge transfer solids (CTS) exhibit a bewildering array of exotic
phases at low temperature, including superconductivity (SC),
SDW, CDW, and bond-order wave (BOW) and (related) 
spin-Peierls (SP) phases. 
Among 
the intriguing parallels between organic and high temperature 
superconductors
is the proximity of the SC phase to the SDW \cite{SC,Jerome}.
Given the enormous focus on understanding the ``normal'' state of high 
temperature superconductors, it seems clear that an analogous effort should be
made to understand the ``normal'' state of organics that exhibit SC. The
SC-SDW proximity then suggests that it is  crucial to understand the 
precise nature of the
SDW and the mechanism of its formation, as well as the roles played by 
dimensionality effects and strong Coulomb interactions
\cite{SC,Jerome}.   
Very recent experiments on several materials indicate that the SDWs in the 
organics are highly
unconventional in nature 
\cite{Pouget1,Pouget2,Toyota1,Kartsovnik,Iye,Caulfield,Miyagawa},
imposing strict requirements of any theory of the
SDWs.

In this Letter, we present a unified theory of SDW formation in a large
class of 2:1 cationic CTS.
While we believe that our theory
is general, we focus on the specific
materials (TMTSF)$_2$PF$_6$, (TMTTF)$_2$Br and (BEDT-TTF)$_2$MHg(SCN)$_4$
(M = K, Rb, Tl), in which the unconventional nature of the SDW 
has been demonstrated recently. In the past, experiments by 
several groups have established that the low temperature insulating phases
in (TMTSF)$_2$PF$_6$ \cite{Wzietek} and (TMTTF)$_2$Br \cite{Torrance} are
SDWs. Surprisingly,
recent X-ray scattering experiments \cite{Pouget1,Pouget2} 
have revealed features associated
with CDW in both materials even for T $<$ T$_{SDW}$.
In (TMTTF)$_2$Br, the experiment finds
signatures of even a 4k$_F$ lattice displacive instability (k$_F$ = 
Fermi wavevector), along with the more usual 2k$_F$ charge instability 
{\it within} the SDW phase \cite{Pouget1,Pouget2}. 
The status of the experiments in (BEDT-TTF)$_2$MHg(SCN)$_4$
is very similar: while early magnetic susceptibility \cite{Toyota} 
and $\mu$sr \cite{Pratt} measurements
provided clear evidence for a SDW, measurement of angle-dependent 
magnetoresistance oscillations \cite{Toyota1,Kartsovnik,Iye,Caulfield} 
have engendered the view that the
insulating phase here is a ``mysterious'' state that is a ``SDW accompanied
by CDW'' or a ``CDW accompanied by SDW'' \cite{Toyota} . 
The authors of a recent
$^{13}$C-NMR study have concluded that the insulating state here 
is not a SDW 
at all, but is a CDW \cite{Miyagawa}.
The authors, however,
also state that they have ``no
idea which kind of CDW reconciles the susceptibility anisotropy ... and other
magnetic properties.''

Coexistence of CDW-SDW {\it of the same periodicity}
is outside the scope of standard theories
of density waves in the organic CTS \cite{Bourbonnais,Gorkov}, 
which emphasize the
nesting associated with the quasi-one dimensional (quasi-1d) Fermi surface
of these materials. Although CDW, SDW and SP instabilities are possible 
within these theories, these instablities occur in {\it nonoverlapping}
parameter
regions, making coexistence impossible.
The microscopic theory we present here explains the
puzzling CDW-SDW coexistence in a natural fashion. The attractive features of
our theory include: (i) our Hamiltonian is the standard strongly
correlated model for quasi-1d organic conductors \cite{Bloch},
and although the manifestations are novel, no exotic interactions are 
necessary to generate them; (ii) the theory can explain the differences
between structurally similar quasi-1d systems
that exhibit the SP phase and quasi-2d systems that exhibit the
SDW;  and finally, (iii) the theory clarifies the limitations
of single-particle nesting concepts \cite{Gorkov} and
of the theoretical bias that SDW exists 
only at or near the 1/2-filled band.

We posit that the Hamiltonian appropriate for the materials considered
here is the quasi-2d extended Hubbard model,
$$H = H_0 +H_{ee} + H_{inter} \eqno(1a)$$
$$H_0 = -\sum_{j,M,\sigma}[t-\alpha(\Delta_{j,M})]B_{j,j+1,M,M,\sigma} 
+ \beta\sum_{j,M}v_{j,M}n_{j,M} $$ 
$$+ K_1/2\sum_{j,M}(\Delta_{j,M})^2 + K_2/2 \sum_{j,M}v_{j,M}^2 \eqno(1b)$$
$$H_{ee}=U\sum_{j,M}n_{j,M,\uparrow}n_{j,M,\downarrow} +
 V\sum_{i,M}n_{j,M}n_{j+1,M} \eqno(1c)$$
$$H_{inter} = -t_{\perp}\sum_{j,M,\sigma}
B_{j,j,M,M+1,\sigma}
\eqno(1d)$$
In the above, $j$ is a site index while $M$ is a chain index,
$B_{j,k,L,M,\sigma} \equiv 
[c_{j,L,\sigma}^\dagger c_{k,M,\sigma} + h.c.]$,
$\Delta_{j,M}=(u_{j,M}-u_{j+1,M})$,
where $u_{j,M}$ is
the displacement of the molecular site from its
equlibrium position,
and $v_{j,M}$ represents an intramolecular
vibration. The total Hamiltonian describes coupled chains, with on-site Coulomb
interaction $U$, intrachain
nearest-neighbor Coulomb interaction $V$, and intra- and
interchain nearest-neighbor hoppings $t_j = t-\alpha(u_{j,M}-u_{j+1,M})$
and $t_{\perp}$. For simplicity, we assume a rectangular lattice.
We are interested in the realistic parameter regime $t_{\perp}$ $\sim$ 0.1,
$V$ $\sim$ 2$|t|$, $U$ $>$ 4$|t|$.

A critical implicit parameter in the Hamiltonian of Eq.~(1) is the bandfilling.
Charge-transfer from the cations to the inorganic
anions in the 2:1 CTS of interest leads to one hole per two organic molecules,
i.e., a 1/4-filled band of holes. This is the bandfilling we consider. 
Based on the weak dimerization along the stack axis
(observed even {\it above} the metal-insulator
transition temperature T$_{MI}$ \cite{SC,Jerome}),
it is commonly argued that 2:1 cationic CTS can be modeled as
{\it effective} quasi-1d 1/2-filled band systems
\cite{Bourbonnais,Gorkov} with Fermi surface nesting.
We show that although some aspects of the physics of
the {\it strongly correlated} 1/4-filled band {\it can} be understood
within the {\it weak coupling} effective 1/2-filled theory, 
others simply cannot.
In particular, since the coexistence of CDW-SDW with the
same periodicity is impossible within the 1/2-filled band \cite{Baeriswyl},
the recent observations \cite{Pouget1,Pouget2,Toyota1,Kartsovnik,Iye,Caulfield,Miyagawa}
clearly preclude this scenario as a 
consistent description of the normal state of these CTS.
We postpone further discussion of this issue until later, when we show that
both from theoretical and experimental perspectives, 
the weak high temperature
dimerization cannot lead to a modulation of $t_j$. 

An intuitive understanding of the broken symmetry coexistence within Eq.~(1)
can be obtained in the 1d limit,   
$t_{\perp}$ = 0. Since long-range SDW does not occur here, 
the relevant order parameters are the site charge
density and the bond-order
$B_{j,j+1,M,M,\sigma}$.
Periodic modulations of the charge density lead to the CDW, and of the
bond-order to the BOW. Previous work has established \cite{Ung2}
that the periodic lattice distortion arising from the BOW has the form
 ${u_j = u_0[r_2cos(2k_Fj - \theta_2) + 
r_4cos(4k_Fj - \theta_4)]}$, where $r_2$ and $r_4$ are the relative weights of
the 2$k_F$ and 4$k_F$ components, and $\theta_2$ and $\theta_4$ are the 
corresponding
phase angles \cite{Ung2}. In contrast, the CDW can have {\it either}
the
2$k_F$ {\it or} the 4$k_F$ modulation {\it but not both}, so  
${n_j = 1/2 + n_0(cosQj - \phi)}$,,
where $Q$ = 2$k_F$ or 4$k_F$ \cite{Ung2,Ung1}. For comparison
with what follows, we sketch in Figs.~1(a) and (b) the familiar BOW
and SDW configurations for the 1/2-filled band.
\begin{figure}[h]
\centerline{\epsfig{bbllx=0pt,bblly=0pt,bburx=618pt,bbury=180pt,
	width=3in,file=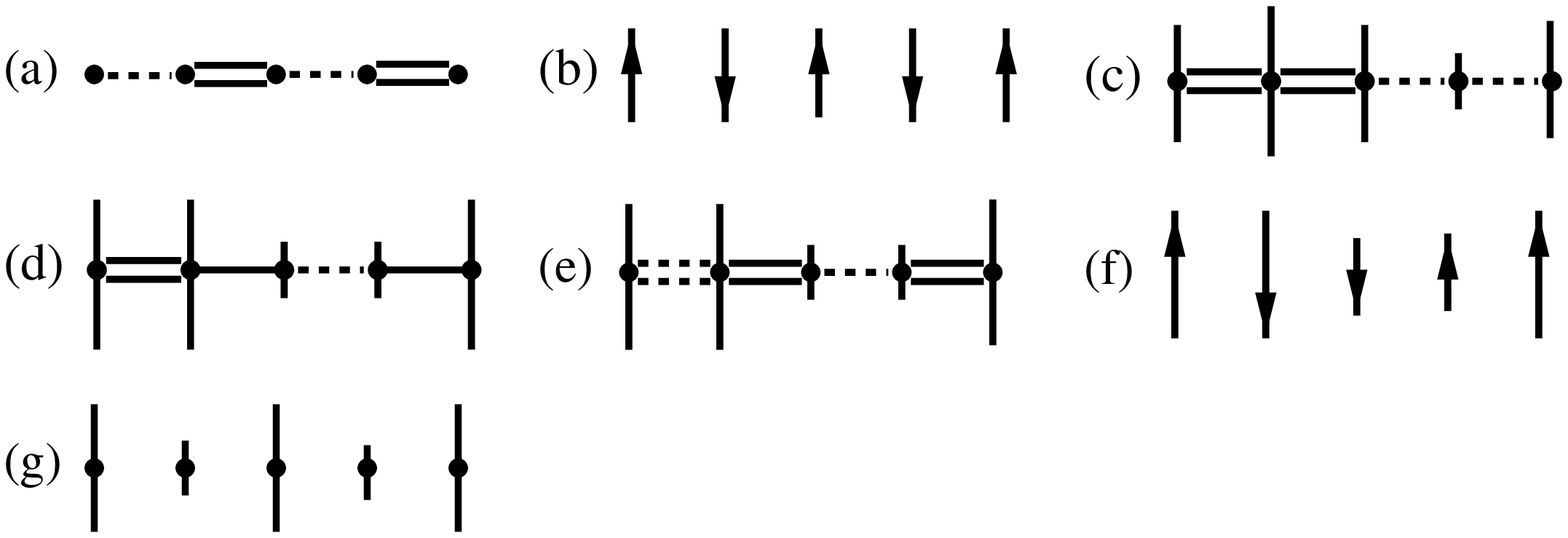}}
\caption{Schematics of the 1d (a) 1/2-filled BOW, (b) 1/2-filled
SDW, (c) 1/4-filled uncorrelated 2k$_F$ BOW-CDW, (d) 1/4-filled
correlated 2k$_F$ BOW-CDW (the 4k$_F$ BOW is the same as (a)),
(e) 1/4-filled BOW-CDW that is a superposition
of 4k$_F$ (Fig.~1(a)) and 2k$_F$ (Fig.~1(d)) periodicities, 
and that occurs for nonzero
$V < V_c$, (f) 1/4-filled 2k$_F$ SDW investigated here, and (g) the
4k$_F$ site-diagonal CDW that occurs only for $V > V_c$.
The double
(dotted) bonds are strong (weak), the single bond is of intermediate
strength, and the double dotted bond in (e) is a weak bond that is
stronger than the single dotted bond. The heights of the vertical bars
(arrows) on sites give their relative charge (spin) densities.}
\end{figure}
A crucial feature of non-1/2-filled commensurate bands is the {\it symbiotic}
coexistence between the BOW and CDW \cite{Ung2,Ung1}; as
shown in these references, this coexistence also enables one to understand
the general case from the $\beta=0$ limit, which we henceforth adopt.
Importantly, the
1/4-filled band is unique among the non-1/2-filled bands in
that Coulomb interactions can drive BOW-CDWs that are {\it different}
from that driven by electron-phonon interaction. For instance,
as the Hubbard $U$ is increased from
zero, the phase angles ($\theta_2$, $\phi$) of the 2$k_F$ BOW-CDW switch from
(0,$\pi$/4) to ($\pi$/4,0) \cite{Ung2}.  
We show the uncorrelated and correlated 2$k_F$
BOW-CDWs in Figs~1(c) and
(d), respectively. For nonzero $V < V_c$ (where $V_c$ = 2$|t|$ for 
${U \to \infty}$, and is larger for finite $U$) the absolute ground state
acquires a 4$k_F$ BOW character [$r_4 \neq$ 0, $\theta_4$ = 0, 
see Fig.~1(e)], 
but the CDW continues to have
periodicity 2$k_F$ \cite{Ung2,note1}.
Our numerical results will establish 
that the BOW-CDWs of Figs.~1(d) and (e) can
coexist with the particular SDW shown in Fig.~1(f), consistent
with the experimental data.

We begin our numerical analysis in the 1d limit. Since a true long-range
SDW cannot occur here, we incorporate an additional
(external field-like) term $H_{SDW}= -\sum_j\epsilon [n_{j,\uparrow} cos(2k_Fj)
+ n_{j,\downarrow} cos(2k_Fj+\psi)]$
and consider the ground state of ${H + H_{SDW}}$.
$H_{SDW}$ imposes a SDW in the 1/2-filled band for $\psi = \pi$ 
and the SDW of Fig.~1(f) in the 1/4-filled band 
for $\psi = \pi/2$, with the amplitude of the SDW increasing with $\epsilon$.
We examine the BOW-SDW coexistence using exact finite size
calculations. Specifically, we calculate the exact electronic ground state
energies
$E(\alpha u_O, \epsilon)$ of 
finite periodic rings as functions of $\epsilon$, where
$\alpha u_0$ is a rigid bond modulation parameter. 
The quantity
$\Delta E(\alpha u_0, \epsilon) = E(\alpha u_0 = 0, \epsilon) - 
E(\alpha u_0 \neq 0, \epsilon)$ is a
direct
measure of the energy gained on bond distortion. 
For the 1/4-filled band,
it is also necessary to specify $r_2/r_4$. While we have confirmed 
that our results are
valid for many different $r_2$ and $r_4$, we show in the following the specific cases
of $r_2$ = 0 (2k$_F$ BOW-CDW only) and $r_2$ = $r_4$ (both 2k$_F$ and
4k$_F$, contributing equally).

In Fig.~2 we show the behavior of 
${\Delta E(\alpha u_0, \epsilon)}$ for  
a 1/2-filled band of 10 sites and a 1/4-filled band of 16 sites,
for $U$ = 6, $V$ = 1. Because of the larger energy gain the 1/2-filled
band, the nonzero values of $\alpha u_0$ are 0.05 (in units of $|t|$)
in the 1/2-filled case and 0.1 in the 1/4-filled case, respectively.
$\Delta E(\alpha u_0, \epsilon)$ decreases rapidly with
$\epsilon$ in the 1/2-filled band, in agreement with the known result that the 
SDW and the
BOW are mutually exclusive here \cite{Baeriswyl}. 
In contrast, we find that 
$\Delta E(\alpha u_0, \epsilon)$ {\it increases} with $\epsilon$ in the 
1/4-filled band
for both the cases studied, indicating a {\it cooperative} interaction 
between the BOW and the
SDW.
 \begin{figure}[h]
\centerline{\epsfig{bbllx=12pt,bblly=32pt,bburx=576pt,bbury=419pt,
	width=3in,file=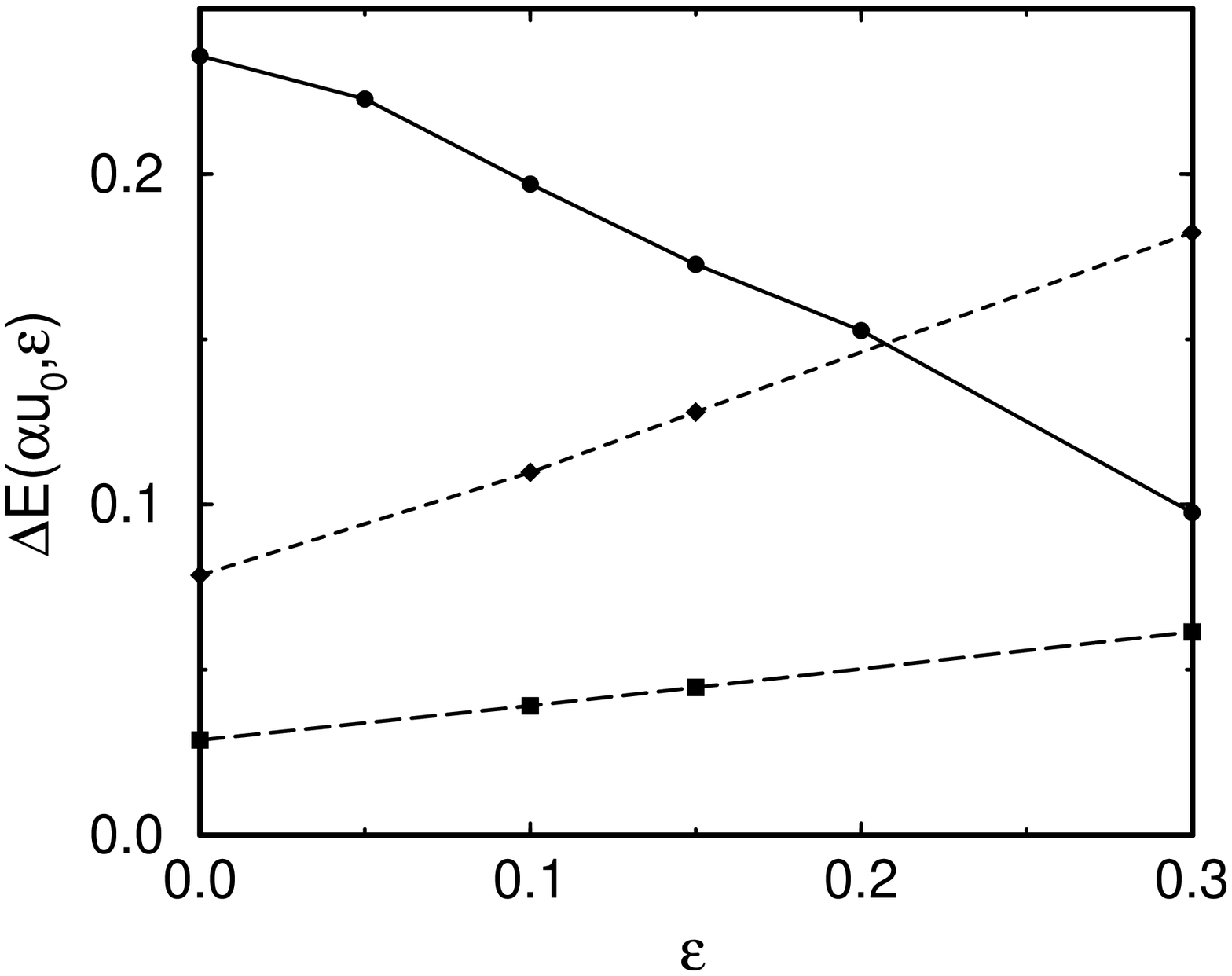}}
\caption{$\Delta E(\alpha u_0, \epsilon)$
versus $\epsilon$ for 
(a) the 1d 1/2-filled band (solid line), (b) the 1d 1/4-filled band with
the bond distortion of Fig.~1(d) (long dashed line), and (c) the
1d 1/4-filled band with the bond distortion of Fig.~1(e) (short dashed line).}
\end{figure}
Since the BOWs we have studied coexist with the correlated 2$k_F$ CDW, the
ground state of ${H + H_{SDW}}$ is an admixture of (2$k_F$ + 4$k_F$)-BOW,
2$k_F$ CDW and 2$k_F$ SDW for the 1/4-filled band. 
\begin{figure}[h]
\centerline{\epsfig{bbllx=162pt,bblly=42pt,bburx=416pt,bbury=470pt,
	width=3in,file=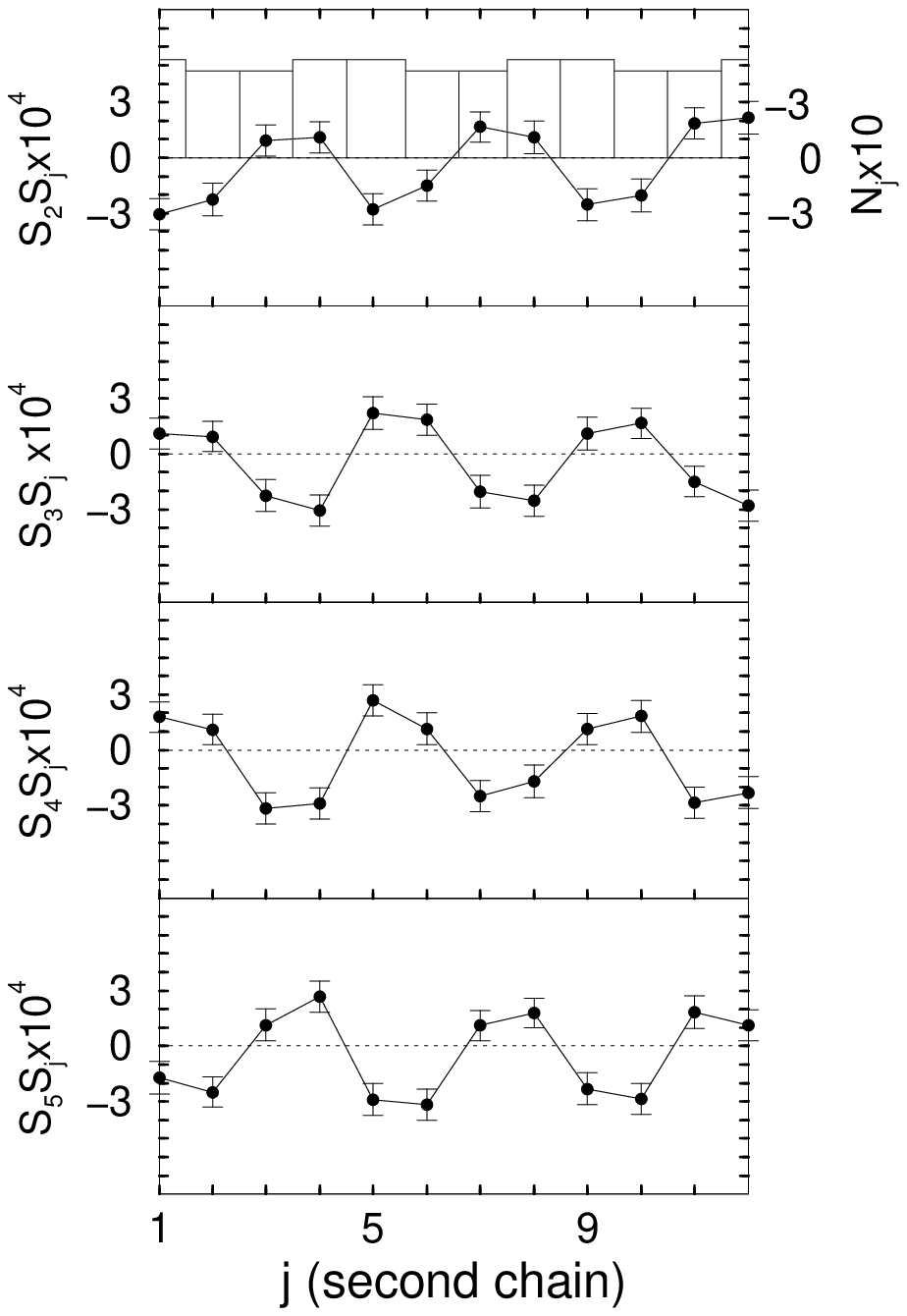}}
\caption{The z-z spin correlations between sites 2,3,4,5 on
the first chain and sites 1 -- 12 on the second chain (see text). The
bars in the top panel show the 
charge densities $N_j$ on the sites of the
second chain.}
\end{figure} 
We have further determined that the bond orders
for nonzero $\epsilon$ 
show a 
modulation of
the type shown in Fig.~1(e) for the 1/4-filled band
{\it even for zero lattice distortion} ($u_0$ = 0),
indicating a tendency for {\it spontaneous} BOW
distortion in the presence of the SDW.

For $H_{SDW}$ = 0 a true SDW
can occur only
for ${t_{\perp} \neq 0}$. We have therefore performed calculations of 
spin-spin correlations, site charge densities and bond-orders
in the ground state of $H$ alone for
coupled chains. 
These
calculations were done using the constrained path quantum Monte Carlo (CPMC)
approach \cite{CPMC}.
In CPMC a constraining wavefunction (in this case the free-electron solution)
is used to approximate nodal boundaries and avoid the sign problem
present in 2D calculations. We expect that sign problems are
less severe for the 1/4 filled band and for the small $t_\perp$
we have used. All calculations were checked against exact results for
a 4 $\times$ 4 lattice. 
The CPMC calculations are for four coupled chains of length twelve 
sites each, with the same values of $U$, $V$, $\alpha u_0$ as in 1d
and $t_{\perp}$ = 0.1.
The 12 $\times$ 4 lattice is taken to be periodic along both directions. 
We incorporate a phase difference of
$\pi$ between the BOWs on neighboring chains, based on calculations 
(a) in the noninteracting limit, and (b) for 
the 4 $\times$ 4
lattice in the interacting cases, that 
indicate that this particular phase difference
gives the lowest total energy \cite{Mazumdar98}. 
A SDW coexisting with the BOW-CDW requires
now antiferromagnetic
{\it interchain} spin-spin correlations. In Fig.~3 we show the spin-spin
correlations between consecutive sites 2,3,4,5 on the first chain and sites
1--12 on the second chain for the case $r_2$ = 0 only.
As seen from the figure,
(a) The 2$k_F$  bond distortion leads to antiferromagnetic
interchain spin-spin correlations,
(b) there is a simultaneous
intrachain antiferromagnetic spin-spin correlation, - the spin
densities on sites 3 and 4 are opposite to those on sites 2 and 5, 
and 
(c) the magnitude of the interchain spin-spin correlation for a given
site on the second chain does not simply decrease with the separation
from the site on the first chain, but is also
determined by the charge density on the particular site on the second 
chain. 
This is a signature
of long range SDW within the distorted lattice. 
With our choice of $t_{\perp}$, the
CPMC technique does not allow us to obtain sufficiently accurate
spin-spin correlations
between sites two chains apart.
However, with a slightly larger $t_{\perp}$
(0.2), these more distant interchain spin-spin correlations are also in
agreement with antiferromagnetic interchain correlations . 
Finally, as in 1d, we have
also performed the calculations for  $r_2$ = $r_4$ 
and find that the
interaction between the BOW-CDW and the SDW remains cooperative, 
decreasing (increasing) $\alpha u_0$ decreases
(increases) the strength of the interchain 
antiferromagnetic correlations \cite{Mazumdar98}.

In Fig.~4 we sketch the ground 
state broken symmetry that emerges from the CPMC results: two adjacent
sites with unequal charge but parallel spins are
surrounded by other such pairs with opposite spins. Viewing the pairs
of sites as a single effective sites, this appears similar to
the SDW of the effective 1/2-filled band
scenario \cite{Bourbonnais,Gorkov}, but there is the critical
distinction that there are different charge and spin densities on the 
individual molecules within the pairs, and this internal structure
will show up in experiments. 
\begin{figure}[h]
\centerline{\epsfig{bbllx=0pt,bblly=0pt,bburx=457pt,bbury=220pt,
	width=2.0in,file=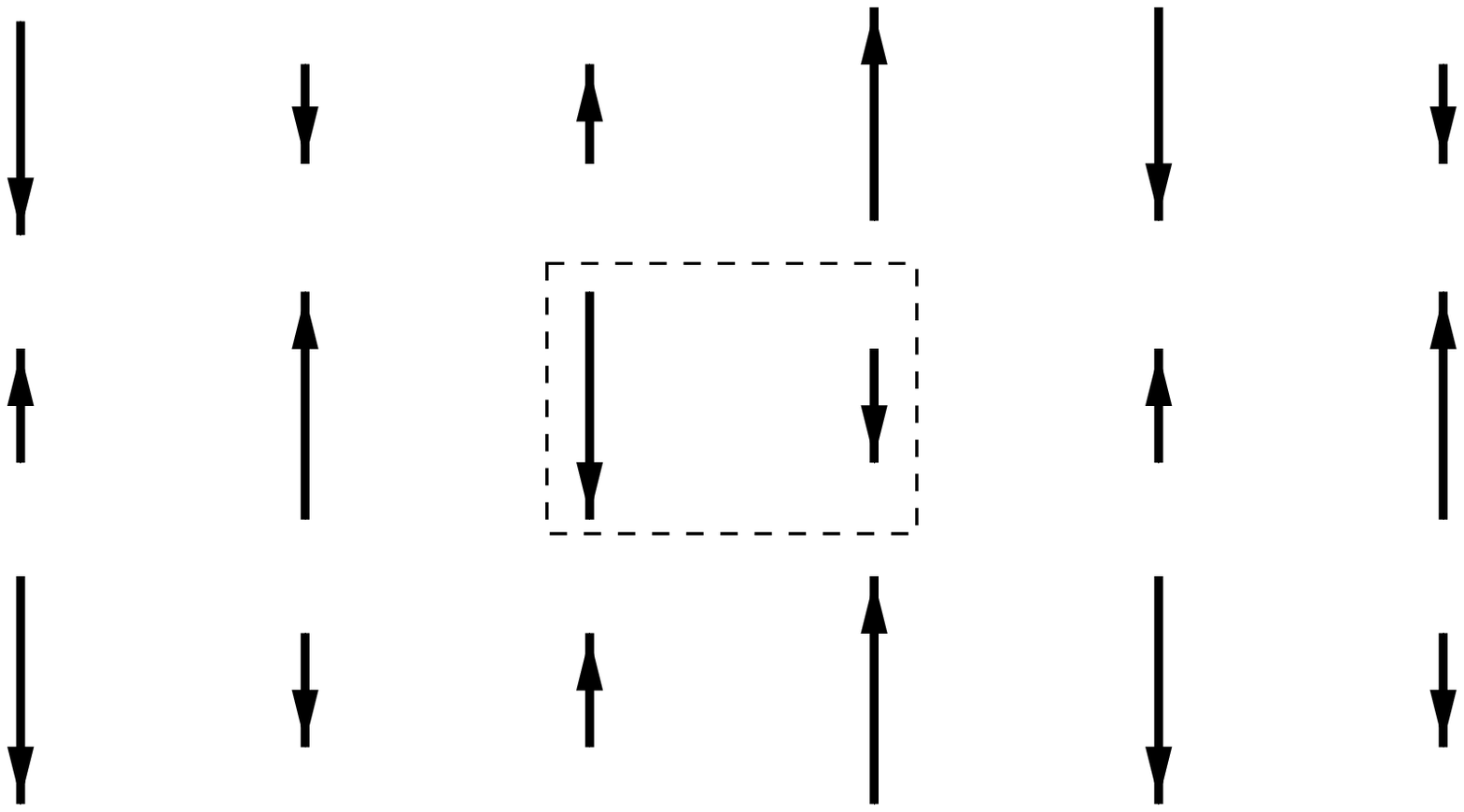}}
\caption{Schematic of the quasi-2d SDW in the correlated
1/4-filled band. The heights of the arrows have the same meaning as in
Fig.~1(f).}
\end{figure}
In their X-ray diffraction
experiment, Pouget and Ravy find strong evidence only for the 2$k_F$ CDW in
the so-called SDW phase of (TMTSF)$_2$PF$_6$, and perhaps both the CDW 
as well as a 2$k_F$ or 4$k_F$
BOW in the SDW phase of (TMTTF)$_2$Br \cite{Pouget1,Pouget2}.
Our results indicate that the CDW-SDW necessarily drives a modulation of the
$t_j$. Whether or not an {\it observable} 
lattice distortion accompanies this merely
depends on the strength of $\alpha$.

We now revisit the issue of the bandfilling. The parameters in Eq.(1)
should be derived from the overall Hamiltonian describing both the organic
cations and the inorganic anions. The crystal structures of (TMTTF)$_2$X and
(TMTSF)$_2$X indicate that the anions
face the ``stronger bonds'' between consecutive
organic molecules \cite{Pouget1}. The anions therefore introduce a 
nonnegligible ``anionic potential,'' the leading term in which
takes the form
${\nu [c_{j,M,\sigma}^\dagger c_{j+1,M,\sigma}+h.c.]}$, where $\nu$
(=1) is the number of (extra) electrons on the anion. This 
interaction
modifies the intrachain $t_j$, {\it decreasing} the
effective hopping integral between the particular pair of organic molecules,
equalizing consecutive $t_j$, and stabilizing a metallic phase.
Very similar 
conclusions have been reached by Fritsch and Ducasse \cite{Fritsch}, who
actually demonstrated the decrease in the hopping integral in question by
taking the anion potential into account. Note that equal effective hopping
integrals at T $>$ T$_{MI}$ is {\it essential} for explaining the 
high temperature metallic conductivity if Coulomb correlations are strong:
any dimerization of the hopping integrals 
leading to an effective
1/2 filled upper subband
should have, in the presence of a nonzero
$U$, opened a Mott-Hubbard gap, leading to an insulator. 

In conclusion, we have shown that the ``normal'' ground state of the 2:1
cationic CTS is a complex mixture of density
waves. In particular, we have discovered 
a {\it cooperative} interaction among the CDW-BOW and 
the SDW, which emerges naturally when
lattice discreteness, Coulomb interactions and
actual bandfilling are taken properly into account. Three final
comments are in order. First, in the appropriate small $t_{\perp}$ regime,
our theory correctly describes the existence of highly 1d systems
(such as (TMTTF)$_2$X, where X$\ne$Br) which exhibit only SP/BOW
and CDW coexistence ({\it i.e.}, no long-range SDW) \cite{Mazumdar98}.
Second, although lack of space precludes detailed
consideration of the important magnetic field-induced SDW phenomena
\cite{Chaikin}, the discussion surrounding
Fig.~4 suggests that key features of the previous approaches \cite{Gorkov}
remain true within the 1/4-filled, strongly correlated framework. In
addition, however,
interesting effects due to unequal charge and spin densities on the
paired sites may emerge. Third, since 
SC appears in these CTS only upon the
melting of the SDW, it seems that any theory of organic SC should take into consideration
the important roles of bandfilling and strong Coulomb interactions
that are established by the present work.

SM acknowledges valuable discussions with L. Ducasse. DKC acknowledges
support from NSF-DMR-97-12765.
RTC acknowledges support of a NSF GRT Fellowship. The numerical 
calculations were done in part at the NCSA.

$^*$ Permanent address: Solid State and Structural Chemistry Unit, Indian
Institute of Science, Bangalore 560012, India.


\begin{references}

\bibitem{SC} {\it Organic Superconductors}, T. Ishiguro and K. Yamaji, Springer
Series Solid State Vol. 88 (Springer Verlag, Berlin, 1990).
\bibitem{Jerome} D. Jerome, Science {\bf 252}, 1509 (1991).
\bibitem{Pouget1}  J.P. Pouget and S. Ravy, J. Phys. I {\bf 6}, 1501 (1996).
\bibitem{Pouget2}  J.P. Pouget and S. Ravy, Synth. Metals {\bf 85}, 1523 (1997).
\bibitem{Toyota1} T. Sasaki and N. Toyota, Phys. Rev. B{\bf 49}, 10120 (1994).
\bibitem{Kartsovnik} M.V. Kartsovnik et al., Synth. Metals, {\bf 70}, 811 (1995)
. 
\bibitem{Iye} Y. Iye et al., J. Phys. Soc. Jpn. {\bf 63}, 674 (1994).
\bibitem{Caulfield} J. Caulfield et al., J. Phys. Cond. Matter
{\bf 6}, L155 (1994).
\bibitem{Miyagawa} K. Miyagawa, A. Kawamoto and K. Kanoda, Phys. Rev. 
B{\bf 56}, R8487 (1997).
\bibitem{Wzietek} P. Wzietek et al., J. Phys. I {\bf 3}, 171 (1993) and references therein.
\bibitem{Torrance}S.S.P. Parkin et al., J. Physique Colloque, {\bf 44}, C-3, 
1111 (1983).
\bibitem{Toyota} T. Sasaki and N. Toyota, Synth. Metals,
{\bf 70}, 849 (1995).
\bibitem{Pratt} F.L. Pratt, T. Sasaki, N. Toyota
and K. Nagamine,  Phys. Rev. Lett. {\bf 74}, 3892 (1995).
\bibitem{Bourbonnais} C. Bourbonnais, Synth. Metals {\bf 84}
, 19 (1997) and
references therein.
\bibitem{Gorkov} L.P. Gor'kov, J. Phys. I {\bf 6},
1697 (1996) and references therein.
\bibitem{Bloch} S. Mazumdar and A.N. Bloch, Phys. Rev. Lett. {\bf 50}, 207
(1983).
\bibitem{Baeriswyl} D. Baeriswyl, D.K. Campbell and S. Mazumdar, in
{\it Conjugated Conducting Polymers}, edited by H. Kiess (Springer-Verlag,
Berlin, 1992).
\bibitem{Ung2} K.C. Ung, S. Mazumdar and D. Toussaint,
Phys. Rev. Lett.
{\bf 73}, 2603 (1994).
\bibitem{Ung1} K.C. Ung, S. Mazumdar and D.K. Campbell, Solid St. Commun., {\bf
85},
917 (1993).
\bibitem{note1} For $V > V_c$, the ground state is a 4k$_F$ CDW
[Fig.~1(g)]. This state has not been observed in this class of CTS.
\bibitem{CPMC} S. Zhang, J. Carlson and J.E. Gubernatis, Phys. Rev. B{\bf 55},
7464 (1997).
\bibitem{Mazumdar98} R. Torsten. Clay, S. Ramasesha, David K. Campbell
and S. Mazumdar, unpublished.
\bibitem{Fritsch} A. Fritsch and L. Ducasse, J. Phys. I, {\bf 1},
855 (1991).
\bibitem{Chaikin} P. Chaikin, J. Phys. I {\bf 6}, 1875 (1996). 
\end{references}
\end{document}